# Lineshape of the thermopower of quantum dots

**S Fahlvik Svensson[1,5], A I Persson[1,2], E A Hoffmann[2,3], N Nakpathomkun[2,4], H A Nilsson[1], H Q Xu[1], L Samuelson[1] and H Linke[1,2,5]**

[1]Solid State Physics and The Nanometer Structure Consortium (nmC@LU), Lund University, Box 118, 221 00 Lund, Sweden

[2]Physics Department and Materials Science Institute, University of Oregon, Eugene, OR 97403, USA

[3]Center for NanoScience and Fakultät für Physik, Ludwig-Maximilians-Universität, 80539 Munich, Germany

[4]Department of Physics, Thammasat University, Pathum Thani, Thailand, 12120

E-mail: sofia.fahlvik_svensson@ftf.lth.se, heiner.linke@ftf.lth.se

**Abstract.** Quantum dots are an important model system for thermoelectric phenomena, and may be used to enhance the thermal-to-electric energy conversion efficiency in functional materials. It is therefore important to obtain a detailed understanding of a quantum-dot's thermopower as a function of the Fermi energy. However, so far it has proven difficult to take effects of co-tunnelling into account in the interpretation of experimental data. Here we show that a single-electron tunnelling model, using knowledge of the dot's electrical conductance which in fact includes all-order co-tunneling effects, predicts the thermopower of quantum dots as a function of the relevant energy scales, in very good agreement with experiment.

## 1. Introduction

The thermovoltage, $V_{th}$, of a quantum dot (QD) is defined as the open-circuit voltage in response to an applied temperature differential, $\Delta T$, and was explored already early [1, 2] as one of a QD's fundamental electron transport characteristics. In recent years, the thermoelectric properties of QDs have been attracting renewed interest because QDs essentially are highly tunable energy filters, which can be used to optimize their performance as thermoelectric power generators or coolers. Specifically, QDs with a very narrow (delta-like) transmission resonance have been shown to be ideal thermoelectric energy converters: they can be operated reversibly, near the fundamental Carnot limit of their efficiency [3-5], and their efficiency at maximum power approaches the fundamental Curzon-Ahlborn limit [6-8]. The use of QDs (nanocrystals) embedded into bulk material [9, 10] or nanowires [5] has been proposed as a way to enhance a material's thermopower or Seebeck coefficient, $S = V_{th} / \Delta T$. Because the thermopower of ideal QDs depends only on fundamental constants, QDs have also been proposed as a quantum measurement standard for the Seebeck coefficient [11, 12].

For each of these applications, as well as for fundamental understanding, it is important to know how $V_{th}$ depends on the properties of the QD, as well as on the chemical potential in the electron reservoirs. For a dot with infinitely sharp, delta-like transmission resonances separated by an energy $\Delta E$ (figure 1), Beenakker and Staring [1] predicted a sawtooth-like dependence of $V_{th}$ on chemical potential, with a resulting maximum thermovoltage $V_{th}^{max} = (\Delta E/4e) \cdot (\Delta T/T)$. This model suggests that $V_{th}$ of a QD can, in principle, be made arbitrarily large by decreasing the dot size, and thus increasing $\Delta E$ through charging or quantum confinement effects. Staring et al [2] did indeed find a QD's thermovoltage to oscillate with a sawtooth-like lineshape for thermal energies $kT$ much less than the QD's charging energy, $e^2/C$, but found a very much lower amplitude than predicted. This difference was suggested to be due to the finite energy-broadening, $\Gamma$, of the levels in the QD (figure 1), a result of co-tunnelling in the transport through the dot, which depends on the strength of the coupling to the leads [13]. Dzurak et al shared this interpretation in the context of a similar experiment [14]. Turek and Matveev [15] developed a thermopower theory including co-tunnelling and predicted a transition, as a function of decreasing temperature, from a sawtooth-like lineshape to a lineshape more similar to an energy-derivative of the dot's conductance. Scheibner et al [16] observed this transition, but it proved difficult to find detailed agreement between experiment and theory.

---

[5] Authors to whom any correspondence should be addressed.

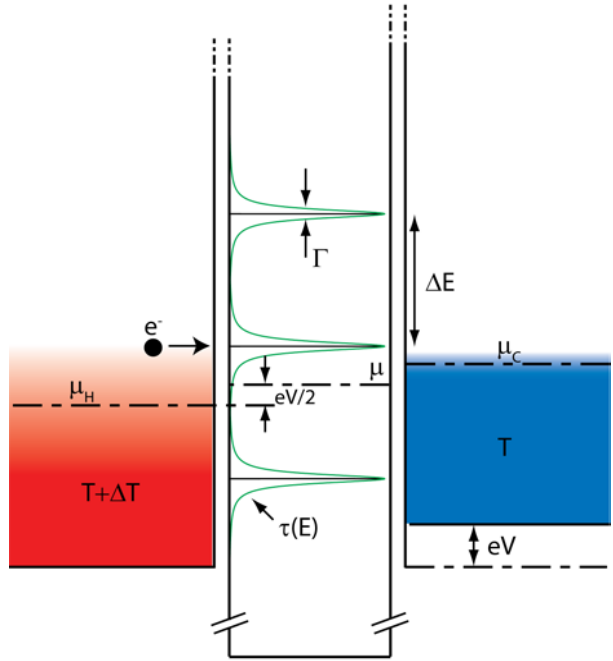

**Figure 1**. We consider a quantum dot (QD) defined by a pair of energy barriers, and attached to a hot (left) and a cold (right) electron reservoir. The QD is characterized by an electron transmission function $\tau(E)$ consisting of Lorentzian resonances with full width at half-maximum $\Gamma$ and energy separation $\Delta E$.

From these prior results it thus emerges that the lineshape of $V_{th}$ of a QD depends on the energy scales $kT$ and $\Delta E$ as well as on $\Gamma$. However, it is known that information about each of these parameters is contained in measurements of the conductance, $G$, of a QD. In this work, we show that, using measurements of $G$ as a function of the chemical potential in the electron reservoirs as an input, and using a simple tunnelling model implemented by the Landauer formalism, we can predict $V_{th}$ as a function of the chemical potential for various $\Delta E/kT$ and $\Gamma/kT$. This approach takes into account transport through all-order tunneling processes simply by using a Lorentzian transmission function with finite broadening $\Gamma$. For comparison to experiment, we use two different QDs defined in heterostructure nanowires, which show a high-quality $S$ signal, and find good qualitative and quantitative agreement with our model. We begin by presenting our modelling results and will then turn to comparison with experiment.

## 2. Model

Within the Landauer formalism the current through a QD connected to a cold (C) and a hot (H) reservoir is given by,

$$I = \frac{2e}{h} \int \left[ f_C(E, \mu_C, T) - f_H(E, \mu_H, T + \Delta T) \right] \tau(E) dE \qquad (1)$$

where $f_{C/H}$ is the Fermi-Dirac distribution on the hot (H) and cold (C) side, respectively, $\mu_{C/H}$ are the chemical potentials in the reservoirs, $T$ and $T + \Delta T$ are the electronic temperatures, and $\tau(E)$ is the transmission function of the QD (figure 1). Here we take only elastic, sequential tunnelling processes into account and no higher order tunnelling processes. We approximate $\tau(E)$ as the sum of Lorentzian functions,

$$\tau(E) = \sum_n A_n \frac{(\Gamma_n / 2)^2}{(E - E_n)^2 + (\Gamma_n / 2)^2} \qquad (2)$$

with their centers located at the dot's resonant levels $E_n$, with widths $\Gamma_n$ and amplitudes $A_n$. For computational reasons it is convenient to define the average chemical potential $\mu = (\mu_C + \mu_H)/2$ and to assume that $V$ is applied symmetrically across the dot such that $\mu_{C/H} = \mu \pm eV/2$. We calculate $S = V_{th}/\Delta T$ using the linear-response approximation [12, 17]

$$S = \frac{V_{th}}{\Delta T} = -\frac{1}{e(T + \Delta T/2)} \frac{\int (E - \mu) \frac{\partial f_0}{\partial E} \tau(E) dE}{\int \frac{\partial f_0}{\partial E} \tau(E) dE}, \qquad (3)$$

which is valid when $eV_{th} << kT$ and $\Delta T << T$. Here, $f_0$ is the equilibrium Fermi-Dirac distribution for $\Delta T = 0$ and $V = 0$. We find that Eq. (3) agrees well with exact numerical solutions of $S$ obtained indirectly by calculating the external bias, $V$, at which $I = 0$ in Eq. (1).

Figure 2(a) shows $V_{th}$ as a function of $\mu$ for a QD with resonant energy levels of width $\Gamma$ varying over a wide range, from $10^{-6} kT$ to $kT$ and for $\Delta E/kT >> 1$. Also plotted in figure 2(a), for comparison, is the ideal sawtooth lineshape predicted by Beenakker and Staring [1] and obtained from Eq. (3) when $\tau(E)$ consists of delta functions ($\Gamma \to 0$). A key observation in figure 2(a) is that the lineshape of $V_{th}(\mu)$ deviates from the sawtooth lineshape for $\Gamma$ as small as $10^{-5} kT$, that is, even for exceedingly small coupling to the leads. With increasing $\Gamma$ the lineshape increasingly resembles the shape of an energy-derivative of the conductance peaks, $V_{th}$ between transmission resonances becomes suppressed, and the amplitude of the $V_{th}$ resonances is, for finite $\Gamma$, much lower than predicted in ref. [1].

The single-electron tunnelling picture combined with the Landauer equation (Eq. 1) provides a conceptual explanation for the observations in figure 2(a). First, we recall that $V_{th}$, for a given configuration ($\Delta T$, $\mu$), can be found by looking for the value of the bias voltage where $I = 0$ (open-circuit condition). When $\tau(E)$ is a delta function, zero current across the dot is obtained when $\tau(E)$ is at the position where the term $\Delta f = f_C - f_H = 0$ (this is the condition of energy-specific equilibrium as defined in ref. [18]). This condition yields the sawtooth lineshape [1]. For finite $\Gamma$, when $\tau(E)$ is slightly broadened, charge carriers can leak through the tails of the Lorentzian. Because $\Delta f$ is not antisymmetric for finite bias voltage, a correction in bias voltage is needed at a given $|\mu - E_n|$ in order to cancel the leakage current and thereby maintaining the zero-current condition. As a result the slope of $V_{\text{th}}(\mu)$ is slightly reduced for small $|\mu - E_n|$ (not visible in figure 2(a)) [12], and falls off for larger $|\mu - E_n|$. However, $\Delta f$ goes to zero faster than a Lorentzian, such that the Lorentzian's tails will always sample $\Delta f$. Therefore, for larger $\Gamma$ and larger $|\mu - E_n|$, the zero-current condition can only be maintained by $V_{\text{th}}$ going to zero.

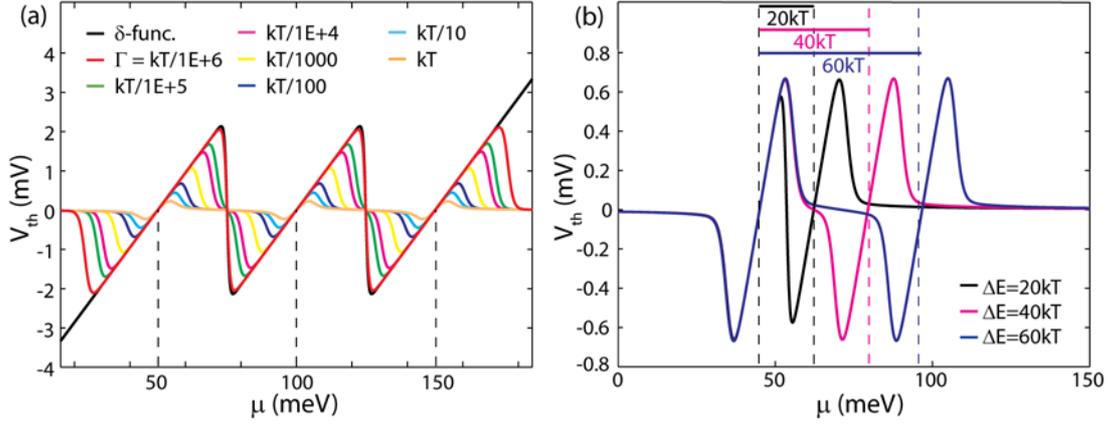

**Figure 2.** Modelling data based on Eq. (3) for $T$ = 10 K and $\Delta T$ = 1 K. (a) $V_{\text{th}}(\mu)$ for varying $\Gamma$ and for $\Delta E$ = 50 meV ≈ 58 $kT$. The dashed lines indicate positions $E_n$ of resonance levels. (b) $V_{\text{th}}(\mu)$ for $\Gamma$ = $kT$/100 and for varying $\Delta E$. The blue graphs in (a) and (b) correspond to approximately equivalent conditions, $\Gamma$ = $kT$/100 and $\Delta E$ ≈ 60 $kT$. The $\mu$-axes use an arbitrary zero.

In figure 2(b) we show $V_{\text{th}}$ as function of $\Delta E$ for fixed $\Gamma$ = 0.01 $kT$. As $\Delta E$ decreases, $V_{\text{th}}$ becomes more triangular, its negative slope between resonances becomes steeper, and its amplitude decreases. It is noteworthy that, for decreasing $\Delta E$, this effect becomes significant already when $\Delta E \sim 10 kT$ and $\Delta E \sim 10^3\Gamma$.

In the above modelling results we have relied entirely on a single-electron tunnelling picture to predict the effects of transport in the presence of weak or strong coupling to the reservoirs. We now turn to experiment to test these predictions.

## 3. Experiment

We used two QDs with different $\Gamma$ and $\Delta E$, each defined by an InP double-barrier structure embedded in an InAs nanowire, using metal-particle seeded growth and chemical beam epitaxy [19, 20]. For sample 1, denoted QD1, InAs$_{0.8}$P$_{0.2}$ was used as the dot material (reducing the effective InP barrier height), and for QD2, InAs was the dot material. Scanning electron microscope inspection after the measurements showed that for QD1 (QD2) the nanowire diameter was 66 nm (55 nm) and the length of the QD (along the wire) was about 190 nm (10 nm).

After growth, the nanowire was transferred onto an n-doped Si wafer with a 100 nm thick SiO$_X$ capping layer. The SiO$_X$/Si substrate acts as a global gate for tuning the QD chemical potential relative to the leads. Ohmic Au/Ni contacts were fabricated by electron-beam lithography, followed by passivation, metallization and lift-off [21, 22]. A three-terminal configuration is used to contact each of the two ends of the nanowire (see inset of figure 3(d)). An AC heating current, $I_H$, applied through terminals A and C, is used to warm the contact electron gas [23]. $I_H$ is applied in a push-pull fashion using a home-built, tunable op-amp circuit such that $I_H$ does not influence the voltage at the nanowire. Terminal B assists in the tuning process. The drain contact temperature at the opposite end of the nanowire increases via electron-phonon coupling in the nanowire [23]. For QD2 we measure $\Delta T$ using quantum-dot thermometry [23]. For QD1 we use finite element modelling [23], which delivers results consistent with those from quantum-dot thermometry, to estimate the effective temperature differential $\Delta T$ applied across the wire at a given cryostat temperature and $I_H$[6].

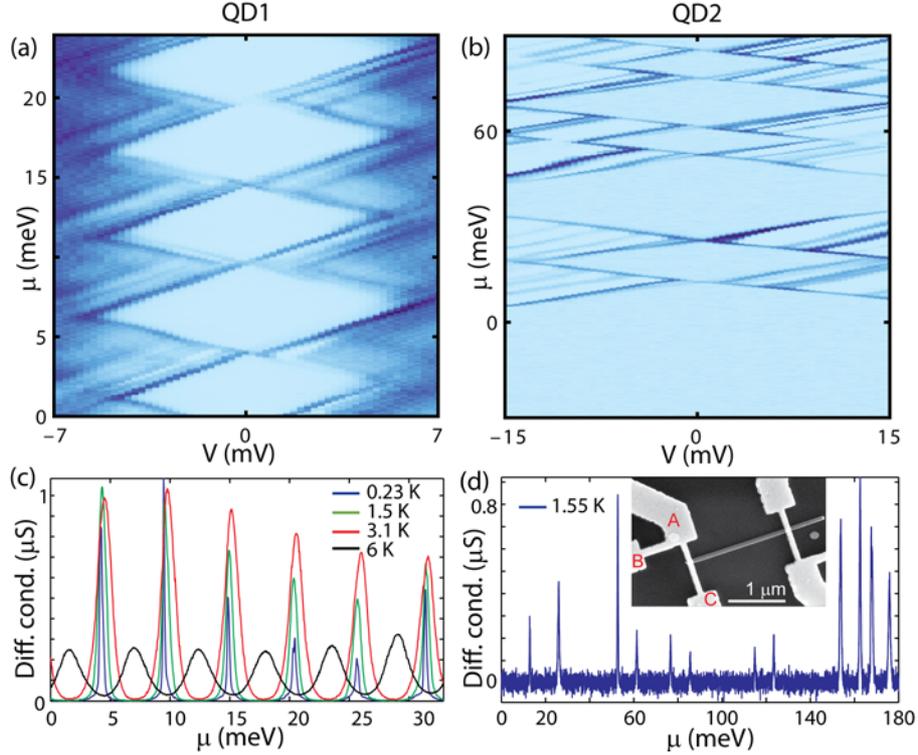

**Figure 3.** Differential conductance as a function of source-drain bias, $V$, and gate voltage, $V_g$, for (a) QD1 at temperature $T = 0.23$ K and (b) QD2 at temperature $T = 1.55$ K, with the vertical axis already converted to $\mu$ (arbitrary zero). Differential conductance peaks measured as a function of $\mu$ at $V = 0$ for (c) QD1 and (d) QD2 are used to extract values for $\Delta E$ and $\Gamma$. The shift of energy levels at $T = 6$ K for QD1 is related to a gate shift. The inset of (d) shows the contact configuration used for both devices.

The heating current, at frequency $f$, generates a thermovoltage, $V_{th}$, composed of two distinguishable signals: (i) an AC thermovoltage at frequency $2f$ (because $V_{th} \propto \Delta T \propto I_H^2$), and (ii), after time-averaging, a DC thermovoltage. For QD1, the AC thermovoltage was measured using a standard lock-in technique to detect the second harmonic signal at frequency $2f$ (here $f = 13$ Hz). The lock-in measurements have the advantage of being relatively fast ($\leq 1$ s per data point) without sacrificing the signal-to-noise ratio. For QD2, the DC component of $V_{th}$ was measured. Any background signal (voltage measured when $I_H = 0$) is subtracted from subsequent raw voltage data measured when $I_H \neq 0$. Additional details are available elsewhere [24]. To isolate the DC signal, a low-pass RC filter damps AC components, and therefore, measurements on QD2 require slow sampling rates (several seconds per data point, using a multimeter with 1 s integration time).

---

[6] The actual electronic temperature differential across the QD is expected to be about a factor 2-3 smaller than $\Delta T$ due to a temperature drop in the InAs leads [23].

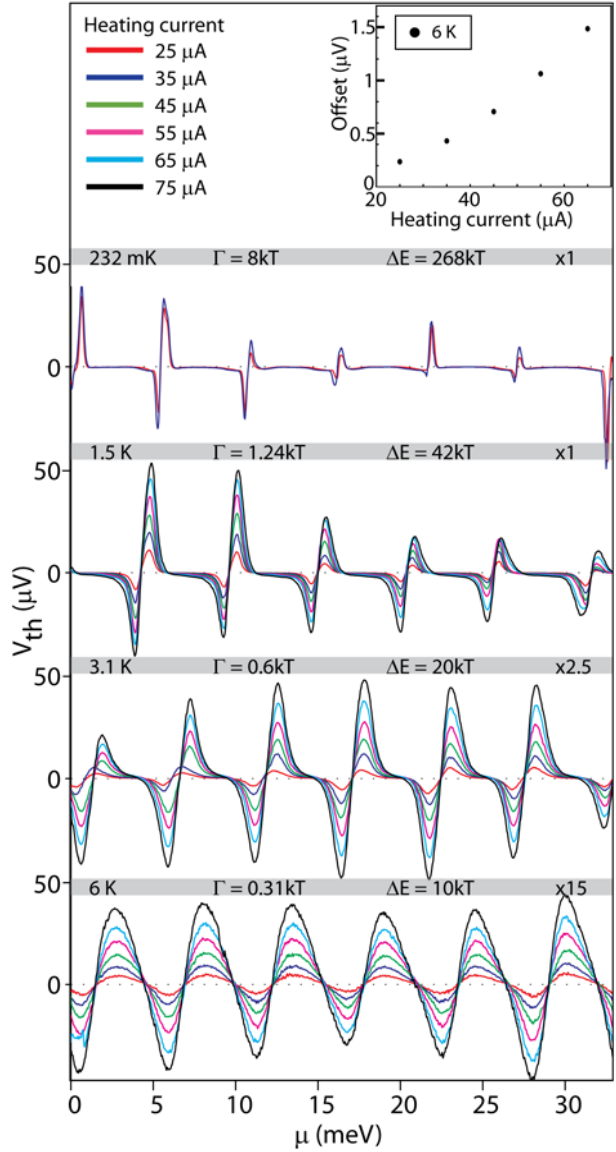

**Figure 4**. Thermovoltage of QD1 as a function of $\mu$ (arbitrary zero) for different $T$ and heating currents $I_H$. Note that the same $I_H$ applied at higher $T$ causes a smaller $\Delta T$. The figure shows the raw data except for the highest temperature at 6K, where an offset (inset) has been subtracted to center the thermovoltage curves at $V_{th} = 0$V (see text). The thermovoltage signals at $T = 3.1$K and 6K were scaled as indicated.

Data that provide a basic characterization of QD1 and QD2 are shown in figure 3. From measurements of the differential conductance, $G = dI/dV$, as a function of gate voltage $V_g$ (see figure 3(a,b)), we convert $V_g$ to an energy scale, $\mu$ [25]. In figure 3(c,d) we show $G$ at $V = 0$ as a function of $\mu$ for both samples. For QD1, the larger of the two dots, conductance peaks are spaced equally indicating that the observed $\Delta E \approx 5.3$ meV is the charging energy $E_C$

= $e^2/C$. By fitting $G$ to the derivative of Eq. 1 at the lowest measured temperature, with $A$ and $\Gamma$ used as free parameters, we find for QD1, averaged over multiple peaks, $\Gamma \approx 160$ µeV, corresponding to 1.8 K. The smaller QD2 has unequally spaced resonances due to quantum confinement effects, with $E_C = 8.7$ meV. Because of a limitation in the temperature range of the cryostat, fitting $G$ of QD1 produces only an approximate upper limit of $\Gamma \approx 30$ µeV, almost an order of magnitude less than in QD1.

In figure 4 we show the thermovoltage of QD1 as a function of $\mu$ for different $I_H$ and cryostat temperature $T$, allowing us to vary the relative energy scales $\Gamma/kT$ and $\Delta E/kT$. Raw data are shown, except for the highest temperature at 6K, where an offset, presumably due to the thermovoltage in the InAs leads[7], has been added to center the thermovoltage curves at $V_{th}$ = 0.

The following observations can be made in figure 4: (*i*) $V_{th}$ increases with $I_H$ as expected in the linear response regime; (*ii*) at the lowest cryostat temperature ($T = 232$ mK), the lineshape of $V_{th}$ is derivative-like as expected from figure 2 for finite $\Gamma$ and $\Delta E >> kT$; (*iii*) with increasing $T$, resulting in a decrease of $\Gamma/kT$ and $\Delta E/kT$, $V_{th}$ becomes more triangular, in agreement with figure 2(b), suggesting that the decrease in $\Delta E/kT$ is the leading cause of the observed lineshape evolution.

## 4. Discussion

After these qualitative observations, we turn to a more quantitative comparison with theory. In figure 5(a) we show $S$ of QD1 at four different temperatures with $I_H$ chosen such that we estimate (based on finite element modelling [23]) a $\Delta T$ of a few 10 mK for each data set. In figure 5(b) we show the corresponding modelling results based on Eq. (3), which used as their only input $T$ and $\Delta T$, as well as $\Gamma = 160$ µeV and $\Delta E = 5.3$ meV as determined experimentally (see figure 3(c)). Agreement between experiment and theory is excellent at the two lowest temperatures, while the suppression of $V_{th}$ between transmission resonances is stronger in the experiment at the higher $T$. $V_{th}$ of QD1 was measured using an AC technique, and therefore the observed suppression might be due to RC damping as the dot's resistance (and therefore the system's RC constant) increases between resonances. Therefore, in figure 5(c), we include the effect of RC damping (at frequency $f = 13$ Hz) using $R(\mu)C = C/G(\mu)$ based on measured data for $G(\mu)$ and the modeled data from figure 5(b). For the capacitance to ground, $C$, we estimate an upper limit of 0.5 nF, with the leading contribution from the cryostat leads ($\approx 0.4$ nF) estimated to be much larger than those from the bond pads and the QD itself. After taking the RC damping into account, the agreement of the model with the experiment is excellent (see figures 5(a) and 5(c)), given the uncertainty in the accuracy of the modelled $\Delta T$.

A similar comparison between experiment and model is shown in figure 5(d)-(f) for QD2, measured using the DC technique, and for smaller values of $\Gamma/kT$ and larger values of $\Delta E/kT$ than in figure 5(a)-(c). Also here experiment (figure 5(d)) and theory (figure 5(e)) agree very well for the lowest temperature. At higher temperatures, the model predicts increasing amplitudes of $S$ due to the decrease in $\Gamma/kT$. However, the same increase is not seen in experiment, where instead $S$ appears to be suppressed compared to the model. To improve the agreement between theory and experiment, we include the effect of the finite load resistance, $R_L = 47$ MΩ, of the DC measurement setup, which is expected to reduce the measured thermovoltage compared to the dot's actual thermovoltage by a factor $(1 + R(\mu, T)/R_L)^{-1}$ (for details, see ref [24]). Because we do not have data of the dot's resistance, $R(\mu, T) = 1/G(\mu, T)$, at all temperatures, we model $G(\mu, T)$ using the equation

$$G(\mu, T) = \frac{1}{kT} G_0 sech^2\left(\frac{\mu - E_n}{2kT}\right), \qquad (4)$$

which is obtained from the derivative of the Landauer equation with respect to bias voltage for $\Gamma \ll kT$ [27], where $G_0$ is a fit parameter that we obtain by matching the measured $G(\mu)$ at $T = 1.55$ K (figure 3(d)), and $E_n$ is the position of the nth resonance as defined in Eq. 2. In figure 5(f), we show the effect of this expected suppression on the modelled thermopower data of figure 5(e), and find that the qualitative agreement with the experiment in figure 5(d) is now excellent. The remaining quantitative disagreement of about a factor of two may be related to an inaccuracy in the value of $\Gamma$ used in the thermovoltage model in figure 5(e). To obtain a more accurate value for $\Gamma$, conductance measurements at lower temperatures $T \approx \Gamma/k$ would be required. Notice that the experimental data also show a curious asymmetric lineshape (of the leftmost resonance) indicating that higher order effects may be important [28].

---

[7] The offset, as shown in the inset to figure 4, increases with heating current. We estimate $\Delta T$ to be about 30-50mK using a heating current of 85 mA, which results in $S \approx 40$ µV/K comparable to the value of 120 µV/K measured for similar InAs nanowires at the higher temperature of 100 K [26].

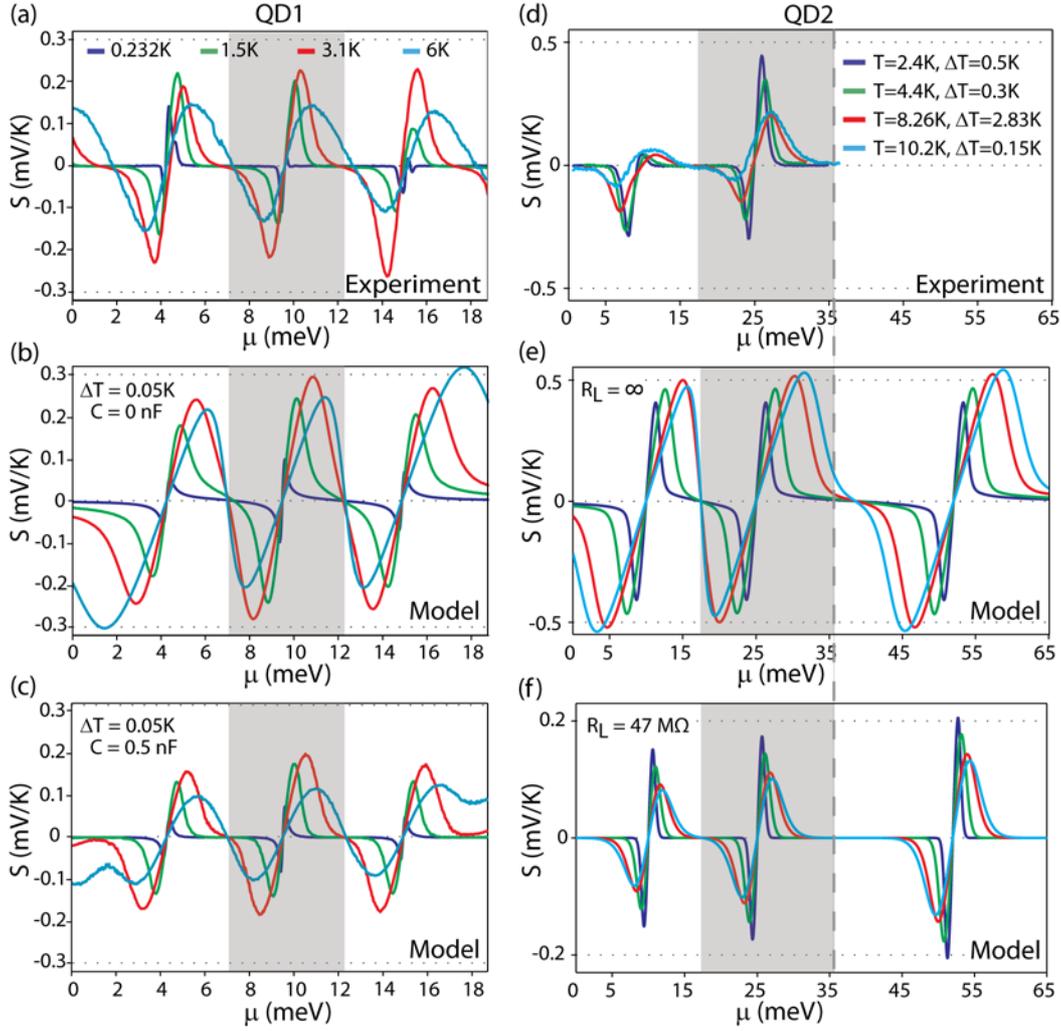

**Figure 5.** (a) Thermopower, $S = V_{th}/\Delta T$, for QD1 at the indicated cryostat temperatures, based on values for $\Delta T$ obtained using finite-element simulations [23]: $\Delta T \approx 0.035$ K at $T = 0.232$ K, $\Delta T \approx 0.05$ K at $T = 1.5$ K, $\Delta T \approx 0.045$ K at $T = 3.1$ K, and $\Delta T \approx 0.01$ K at $T = 6$ K. (b) Modelling results based on Eq. (3) for the same temperatures $T$ as in (a), $\Delta T = 0.05$ K , $\Delta E = 5.3$ meV and $\Gamma = 160$ µeV. (c) The same data as in (b), but including the effect of RC damping in the measurement circuit, using the measured $R = 1/G$ and a fixed $C = 0.5$ nF. Note that the model used only three resonances, and therefore only the center resonance (highlighted by a gray background) experiences the effect of neighboring resonances present in the experiment and corresponds to the data in (a). (d) Thermopower, $S$, of QD2 at cryostat temperatures $T$ and $\Delta T$ as indicated. Here $\Delta T$ was measured using quantum-dot thermometry [23]. $\Gamma/kT$ varies here from 0.145 at 2.4K to 0.042 at 8.26K, and $\Delta E/kT$ from 73 at 2.4 K to 21 at 8.26K. (e) Modelled data based on Eq. (3) for the same $T$, $\Delta T$, $\Delta E$ and $\Gamma_n$ as in (d). Note that the model took into account a third resonance that was present in the device. Only the center resonance in (e) can be compared to (d). (f) The same data as in (e), including the effect of the finite load resistance, $R_L$, of the experimental setup.

In conclusion, we have tested the extent to which a non-interacting, single-particle Landauer model predicts a quantum dot's $S$ lineshape as a function of the relevant energy scales $\Gamma/kT$ and $\Delta E/kT$, which are readily available from conductance measurements. Using high-quality experimental data that allow for a detailed comparison with theory, we find excellent qualitative agreement, and very good quantitative agreement, when effects of the measurement setup are taken into account in the model.

## Acknowledgments

This work was supported by the Swedish Energy Agency *Energimyndigheten* (Grant No. 32920–1), the Office of Naval Research (Grant No. N00014-05-1-0903), an NSF-IGERT Fellowship, the Royal Physiographic Society in Lund, the Sweden-America Foundation, the Thai Government, the Swedish Foundation for Strategic Research (SSF), the Swedish Research Council (VR), the Knut and Alice Wallenberg Foundation and the Nanometer Structure Consortium at Lund University (nmC@LU).